\documentclass{PoS}

\title{$\pi^0$ to two-photon decay in lattice QCD}

\ShortTitle{$\pi^0$ to two-photon decay in lattice QCD}

\author{\speaker{E. Shintani}$^1$\thanks{shintani@het.phys.sci.osaka-u.ac.jp}, S. Aoki$^{2,3}$,
        S. Hashimoto$^{4,5}$, T. Onogi$^1$, N. Yamada$^{4,5}$ (for JLQCD Collaboration)\\
        $^1$Department of Physics, Osaka University, Toyonaka, Osaka 560-0043 Japan,\\
        $^2$Graduate School of Pure and Applied Sciences, 
            University of Tsukuba, Tsukuba 305-8571, Japan, \\
        $^3$Riken BNL Research Center, Brookhaven National Laboratory, Upton, NY 11973, USA,\\
        $^4$KEK Theory Center, Institute of Particle and Nuclear Studies, 
            HIgh Energy Accelerator Research Organization (KEK), Tsukuba 305-0801, Japan,\\
        $^5$School of High Energy Accelerator Science, 
            The Graduate University for Advanced Studies (Sokendai), Tsukuba 305-0801, Japan
}

\abstract{
We calculate the neutral pion ($\pi^0$) to off-shell two photon ($\gamma^*\gamma^*$) 
transition form factor in lattice QCD.
The transition form factor can be extracted from the
three-point function of the form (axial-vector)-(vector)-(vector)
as a function of off-shell two-photon momentum.
Since the axial-anomaly plays an important role in the
$\pi^0\rightarrow\gamma\gamma$ decay process, we employ the overlap fermion, 
which preserves the exact chiral symmetery on the lattice. 
After extrapolating to the chiral and the vanishing photon momentum limit
with a fit function based on vector meson dominance (VMD)
model, we find that  the Adler-Bell-Jackiw anomaly is correctly reproduced.
}

\FullConference{The XXVII International Symposium on Lattice Field Theory\\
                July 26-31, 2009\\
                Peking University, Beijing, China}

\begin{document}

\section{Introduction}

In this report we present our lattice study of the 
$\pi^0\rightarrow\gamma^{(*)}\gamma^{(*)}$ form factor. 
The photons could be either on-shell or off-shell.
The goal of this study is two-fold. 
The first is to reproduce the contribution of the Adler-Bell-Jackiw (ABJ) anomaly~\cite{Adler:1969} 
to the form factor, which is necessary condition for futher applications.
The second is an application to the estimate of the hadronic light-by-light (L-by-L) scattering amplitude, 
which is important in the theoretical calculation of the muon $g-2$, $a_\mu^{\mathrm{L-by-L}}$ \cite{Jegerlehner:2009}.
Since, in contrast to the vacuum polarization diagram, no experimental result 
is avaiable to estimate the L-by-L diagram, purely theoretical estimate is required, 
but it is challenging~\cite{Hayakawa:2005eq}.
Some of the phenomenological estimates is based on the assumption that 
the L-by-L scattering amplitude is dominated by the effective diagram containing 
$\gamma\gamma^*\rightarrow\pi^0$ and $\pi^0\rightarrow\gamma^*\gamma^*$ 
transition form factors connected by a pion propagator.
The results are in the range $a_\mu^{\mathrm{L-by-L}} = (80\sim 130)\times 10^{-11}$
~\cite{Bijinens:1995, Bijinens:1996, Hayakawa:1996, Knecht:2002, Melnikov:2004}.
Even within this assumption, these are significant model dependence.
Therefore, model independent calculation of $\pi^0\rightarrow\gamma^*\gamma^*$ transition form factor 
is a valuable step toward the precise test of the Standard Model~\cite{Olive:2009}.

We study the $\pi^0\rightarrow\gamma^*\gamma^*$ transition
form factor in soft Euclidean momentum region.
Similar attempt is seen in~\cite{Cohen:2008}, 
in which the technique of~\cite{Dudek:2006} for charmonium decay to two photon
state is applied. Since the axial-anomaly plays a key role, 
the exact chiral symmetry and flavor SU(2) symmetry
are desirable in the lattice calculation. We therefore apply 
the overlap fermion formulation.
We use the $N_f=2$ dynamical overlap fermion configurations generated
on a $16^3\times32$ lattice at $\beta=2.3$~\cite{Aoki:2008tq}, which corresponds to the inverse
lattice spacing $a^{-1}=1.67$ GeV. 
Topological charge is fixed to $Q=0$, which induces finite size effect~\cite{Aoki:2007ka}. 
For chiral extrapolation, we take four values of quark mass,
$m_q=0.015,\,0.025,\,0.035,\,0.05$, approximately covering the range of $m_s/6$ to $m_s/2$ 
with $m_s$ the physical strange quark mass.

\section{$\pi^0\rightarrow\gamma\gamma$ transition form factor}

In the continuum theory the $\pi^0\rightarrow\gamma\gamma$ transition
form factor, $f_{\pi^0\gamma\gamma}(p_1,p_2)$, is defined through a matrix element 
of two electromagnetic (EM) currents $V^{\rm EM}$ between the pion state and the vacuum,  
\begin{equation}
  \int d^4x e^{ip_2x}\langle \pi^0(q)| V_\nu^{\rm EM}(x)V_\mu^{\rm EM}(0) |0\rangle =
  \varepsilon_{\mu\nu\alpha\beta}p_1^\alpha p_2^{\beta}
  f_{\pi^0\gamma\gamma}(p_1,p_2), 
\end{equation}
where $p_1,\ p_2$  are Minkowski photon four-momenta and $q=-p_1-p_2$ is the pion
four-momentum. $f_{\pi^0\gamma\gamma}(p_1,p_2)$ can be extracted from the three
point function of an axial-vector current and two EM currents via
\begin{eqnarray}
&&\int d^4x\int d^4y e^{-i(qx+p_1y)} \langle 0|T\{\partial A^3(x)
V_\nu^{\rm EM}(y)V_\mu^{\rm EM}(0)\}|0\rangle \nonumber\\
&&= -\frac{f_\pi m_\pi^2}{-q^2+m_\pi^2}\varepsilon_{\mu\nu\alpha\beta}p_1^\alpha
p_2^{\beta}   f_{\pi^0\gamma\gamma}(p_1,p_2) + \cdots. 
\label{eq:3pt}
\end{eqnarray}
with the pion decay constant $f_\pi=131$ MeV. The ellipsis represents
excited state contributions which have the same quantum number with a pion.
In the Euclidean lattice calculation, only negative $q^{2}$ region is accessible, 
and we mainly focus on the small negative $q^{2}$ region.
Thus, in the following preliminary analysis, we assume that the first term of (\ref{eq:3pt}) 
gives the dominant contribution, and the excited state contributions are ignored.

To evaluate the three-point function on the lattice, we use a combination of
the conserved axial-vector current ($A_\mu^{\rm cv}$) and the non-conserved local EM currents ($V_\mu^{\rm EM\,loc}$).
With the overlap fermion formulation the conserved current takes a non-local form 
$A^{{\rm cv}\,a}_\mu(x) = \sum_{y,z}\bar q(y)\tau^a K_\mu^A(y,z;x) q(z)$ where
$K_\mu^A(y,z;x)$ is a kernel derived from the flavor non-singlet axial transformation~\cite{Kikukawa:1999}.
$\tau^a$ is the generator of flavor SU(2) Lie group normalized by ${\rm tr}\tau^a\tau^b=\delta^{ab}$. 
The local EM current is defined as $V_\mu^{\rm loc\,EM}(x)=Z\bar q(x) Q_e\gamma_\mu q(x)$ 
with a quark charge matrix $Q_e={\rm diag}(2/3,-1/3)$.
The renormalization factor $Z=1.3842(3)$ is obtained non-perturbatively~\cite{Noaki:2008iy}. 
This local current does not satisfy the charge conservation
and thus unphysical contamination (lattice artifacts) can appear 
in the  result as will be discussed later. 
Since the EM current consists of the flavor triplet and singlet parts, the evaluation of the three point function 
$\langle A^{\rm cv\,3}_\rho V_\nu^{\rm loc\,EM}V_\mu^{\rm loc\,EM}\rangle$ requires to 
calculate both connected and disconnected quark diagrams.
In the following analysis, we assume that the disconnected contribution is negligible, which is valid 
in the flavor SU(3) limit.


We define two three-point functions of type (axial-vector)-(EM vector)-(EM-vector) and
(pseudo-scalar)-(EM vector)-(EM vector) by
\begin{eqnarray}
&&G^{AVV}_{\mu\nu}(P_2,Q) = \sum_{x,y} e^{-iQx-iP_2y}
  \Big\langle\nabla A^{\rm cv\,3}(x)V^{\rm loc\,EM}_\nu(y)V^{\rm loc\,EM}_\mu(0)\Big\rangle_c\nonumber\\
&&={\rm tr}[\tau^{3}Q_{e}^{2}]\Big\langle\sum_{x,y,y',z'}e^{-iQx-iP_2y}
  2{\rm Re}\,{\rm Tr}[S_{q}(0,y')\partial^xK^A(y',z'|x)S_{q}(z',y)\gamma_\nu S_{q}(y,0)\gamma_\mu]
  \Big\rangle,\\ 
&& G^{PVV}_{\mu\nu}(P_2,Q) = \sum_{x,y} e^{-iQx-iP_2y}
  \Big\langle 2m_qP^{\rm rot\,3}(x)V^{\rm loc\,EM}_\nu(y)V^{\rm loc\,EM}_\mu(0)\Big\rangle_c\nonumber\\
&&=2m_q{\rm tr}[\tau^{3}Q_{e}^{2}]\Big\langle\sum_{x,y,y'}e^{-iQx-iP_2y}
  2{\rm Re}\,{\rm Tr}[S_{q}(0,x)\Gamma_P(x,y')S_{q}(y',y)\gamma_\nu S_{q}(y,0)\gamma_\mu]\Big\rangle,  
\label{eq:rhs}
\end{eqnarray}
where $\langle\rangle_c$ denotes the connected contraction, while $\langle\rangle$ denotes the 
statistical average, $S_{q}(x,y)$ denotes the quark propagator and $\Gamma_{P}(x,y)=(1-D_{ov}/M_0)\gamma_5(x,y)$.
``tr'' represents the trace over flavor indices and ``Tr'' represents 
the trace over color and spinor indices.
The $P_{1,2}$ and $Q$ are Euclidean momenta of two photons and a pion, respectively.
On the lattice they are dicretized as $2\pi n_\mu/(L_\mu a)$,
with the lattice spacing $a$ and the lattice size $L_{x,y,z}=16,\, L_t=32$.
The pseudo-scalar density operator
$P^{\rm rot\,3}(x)=\bar q(x)\tau^3\gamma_5[(1-D_{ov}/M_0)q](x)$ forms
the axial-Ward-Takahashi identity with $A_\mu^{\rm cv\,3}$.
The reason for considering (\ref{eq:rhs}) will become clear soon.
In the above equation $\nabla A^{\rm cv}$ denotes the forward
derivative $\nabla A^{\rm cv}(x) \equiv \sum_\rho [A^{\rm
cv}_\rho(x+\hat\rho)-A^{\rm cv}_\rho(x)]$. 
We apply the source method for spatial integral over $y$ with $P_2=(0,0,0,P_{2\,t})$ and $\nu=1$. 
After making Fourier transformation for $t_y$ and $x$, we obtain $G^{AVV}_{\mu\nu}(P_2,Q)$ and
$G^{PVV}_{\mu\nu}(P_2,Q)$ for each $\mu=1\sim 4$ and $P_{2\,t}$.  
The remaining momentum $P_1$ is determined by the momentum conservation, $P_1=-Q-P_2$. 

The flavor non-singlet axial Ward-Takahashi (AWT) identity for the three point functions can be expressed as  
\begin{eqnarray}
G^{AVV}_{\mu\nu}(P_2,Q) &=& G^{PVV}_{\mu\nu}(P_2,Q) +
\langle(\delta^3_A V_\nu^{\rm loc\,EM})V_\mu^{\rm loc\,EM}\rangle_c(P_2)\delta^4(Q)\nonumber\\ 
&+& \langle V_\nu^{\rm loc\,EM}(\delta^3_A V_\mu^{\rm loc\,EM})\rangle_c(P_2)\delta^4(Q)
 +  \cdots
\label{eq:AVV}
\end{eqnarray}
where the second and third terms are contact terms coming from lattice
axial transformation, $\delta_A^a V_\mu^b=\bar q\gamma_5\gamma_\mu\tau^a\tau^b q + \bar
q\gamma_\mu\tau^b\tau^a\gamma_5(1-D_{ov}/M_0)q$ for $\langle V_\nu^{\rm loc\,EM}V_\mu^{\rm loc\,EM}\rangle$.
These terms do not contribute to $f_{\pi^0\gamma\gamma}(P_1,P_2)$ as $Q=0$ implies $P_1=-P_2$, for which
the three point function (\ref{eq:3pt}) vanishes due to its Lorentz structure.
Since we ignore the disconnected diagram, there may be some contributions denoted by ellipsis.
Thus, we calculate the both sides independently to estimate the potential size of disconnected diagrams.

The $\pi^0\rightarrow\gamma\gamma$ transition form factor
is extracted from the three-point function with the following momentum assignments:
\begin{eqnarray}
 P_{1} = (0,P_{1\,y},0,0)\textrm{ at $\mu=3$}, & \textrm{  or  }& 
P_{1} = (0,0,P_{1\,z},0)\textrm{ at $\mu=2$}.
  \label{eq:P_2}
\end{eqnarray}
Then the only nonzero term is the one proportional to $P_{1\,\alpha}P_{2\,\beta}\varepsilon_{\mu\nu\alpha\beta}$.
Furthermore, these momentum assignments suppress a class of lattice artifact due to the violation 
of Lorentz symmetry and EM charge conservation.
Since $P_1$ and $P_2$ are orthogonal to each other and have only one
nonzero component, Lorentz violating terms such as
$(P_{1\,\mu}^2+P_{2\,\nu}^2)P_{1\,\alpha}P_{2\,\beta}\varepsilon_{\mu\nu\alpha\beta}$
vanish. Therefore we obtain 
\begin{eqnarray}
  F^{\rm lat}(P_1,P_2) &=& G^{AVV}_{\mu\nu}(P_2,Q)\Big/\Big(\sum_{\alpha\beta}
  P_{1\,\alpha}P_{2\,\beta}\varepsilon_{\mu\nu\alpha\beta}\Big)
= -\frac{f_{\pi} m_\pi^2}{Q^2+m_\pi^2} f_{\pi^0\gamma^*\gamma^*}(P_1,P_2).
\label{eq:F^lat}
\end{eqnarray}
up to contributions from excited states and disconnected diagrams.
Since $P_{1,2}$ are defined in the Euclidean space-time,
the on-shell condition is realized only at zero momentum $P_1=P_2=0$.
Note that we numerically check the consistency of $F^{\rm lat}$ obtained from 
$G^{AVV}_{\mu\nu}$ and $G^{PVV}_{\mu\nu}$ within the statistical error.
$F^{\rm lat}(P_1,P_2)$ from (\ref{eq:F^lat}) are averaged over the 
two physically equivalent momentum assignments in (\ref{eq:P_2}).

\section{Results}

\begin{figure}
\begin{center}
  \vskip 5mm
  \includegraphics[width=85mm]{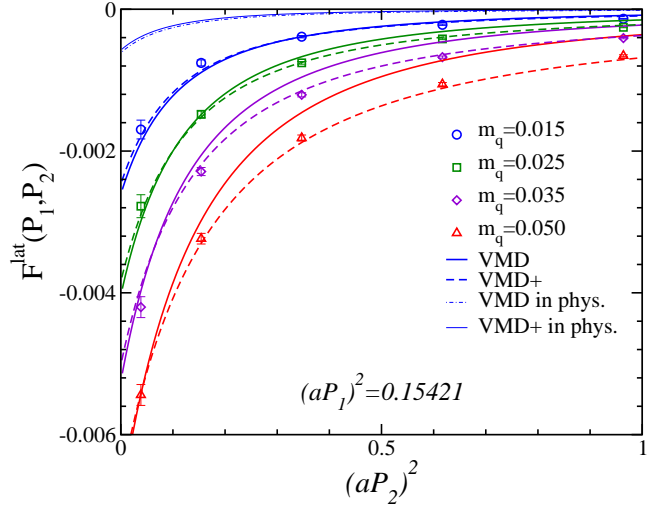}
  \caption{ Lattice results for $F^{\rm lat}(P_1,P_2)$ at a fixed
  $(aP_1)^2\simeq 0.154$ as a function of $(aP_2)^2$ and fit function of 
  $F^{\rm VMD}(P_1,P_2;X_{\rm a})$ (solid curve) and $F^{\rm VMD+}(P_1,P_2;X_{\rm a},c_3,c_4)$ 
  (dashed curve) for each quark mass.
  Thin lines denote the VMD (dashed-dotted) and VMD+(solid) in physical point. } 
  \label{fig:F^lat}
\end{center}
\end{figure}

First we test if the ABJ anomaly~\cite{Adler:1969}
\begin{equation}
  f_{\pi^0\gamma\gamma}(0,0) = \frac{1}{4\pi^2 f_\pi},
  \label{eq:f_p0}
\end{equation}
is reproduced in the chiral and the photon's on-shell limit.
The data are extrapolated to the zero momentum and zero quark mass limit assuming a fit function.
Here we consider the vector meson dominance (VMD) model as an ansatz.
Since we neglected disconnected diagrams, the EM vector current only couples to $\rho$ 
meson.
%
%
The functional form is 
\begin{eqnarray}
&&F^{\rm VMD}(P_1,P_2;X_{\rm a}) = -\frac{m_\pi^2}{Q^2+m_\pi^2} X_{\rm a}G_v(P_1,m_v)G_v(P_2,m_v),
\label{eq:F^VMD}
\end{eqnarray}
with a free parameter $X_{\rm a}$, which is supposed to correspond to the axial-anomaly, $X_a=f_\pi f_{\pi^o\gamma\gamma}(0,0)$.
The vector meson propagator is
\begin{equation}
  G_v(P,m_v) = \frac{m_v^2}{P^2+m_v^2},
\end{equation}
where $m_v$ denotes a vector meson mass.
The vector meson mass is determined independently from an exponential fit of a smear-local
vector current correlator.
Performing chiral extrapolation with a linear function, $m_v=m_v^0+c_vm_\pi^2$, 
we obtain $m_v^0=0.798(19)$ GeV,
which reasonably agrees with the physical $\rho$ meson mass 0.775 GeV. 
Since we encoded the quark mass dependence of $F^{\rm VMD}(P_1,P_2;X_{\rm a})$ into $m_\pi^2$ and $m_v$,
only one free parameter $X_{\rm a}$ is left for the fit of the lattice data.
In Fig.~\ref{fig:F^lat}, we plot $F^{\rm lat}(P_1,P_2)$ at $(aP_1)^2=0.154$, 
which is the minimum non-zero value, as a function of $(aP_{2})^{2}$.
The fit results at different quark masses are also shown by solid curves, 
where the fit point is taken to be $(aQ)^2=0.039$. 
From this plot we can see that lattice data and $F^{\rm VMD}(P_1,P_2;X_{\rm a})$ are in
good agreement at the lowest momentum. 
We obtain
\begin{equation}
  X_{\rm a} = 0.0260(6),
\end{equation}
which is consistent with the expectation $X_{\rm a}=1/(4\pi^2)=0.02533$ despite of various approximations.
Probably the main reason is the exact chiral symmetry, because the existence of 
the conserved axial-vector current significantly reduces lattice artifact.

Going beyond the minimum value of the momenta, the fit curves deviate from the data point.
To accommodate this deviation, we modify the fit form by incorporating an excited vector meson state as
\begin{figure}
\begin{center}
  \includegraphics[width=30mm]{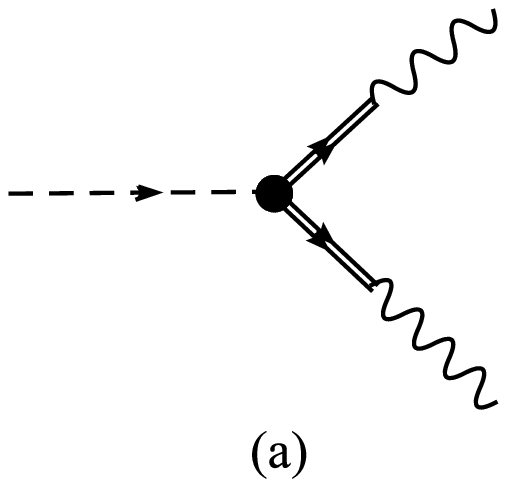}
  \hspace{3mm}
  \includegraphics[width=30mm]{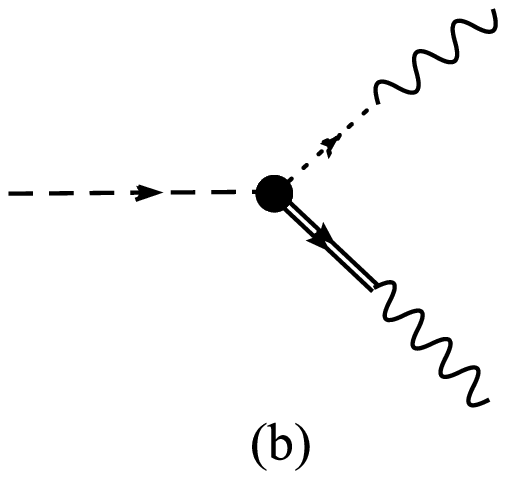}
  \hspace{3mm}
  \includegraphics[width=30mm]{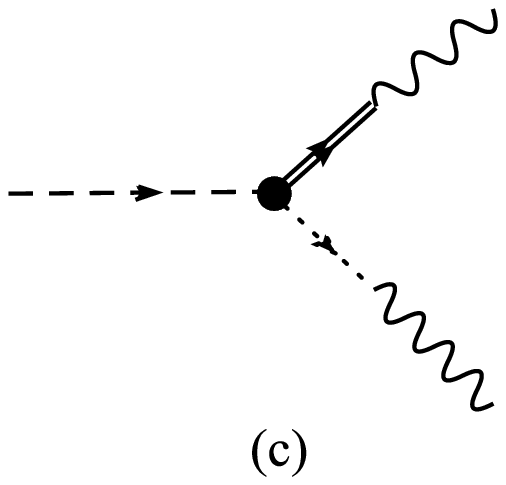}
  \hspace{3mm}
  \includegraphics[width=30mm]{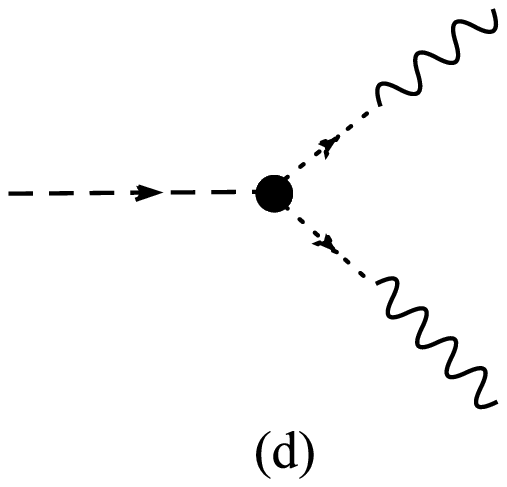}
  \caption{Vertecies of $\pi^0\rightarrow\gamma^*\gamma^*$ decay. The dashed and waved lines show 
  pion and photon propagator, and double-line and dotted lines show vector resonance and its excited state
  propagator respectively. }
  \label{diagram2}
\end{center}
\end{figure}
\begin{eqnarray}
F^{\rm VMD+}(P_1,P_2;X_{\rm a},c_{3,4}) &=& -\frac{m_\pi^2}{Q^2+m_\pi^2} X_{\rm a}\Big[ 
  c_3 G_v(P_1,m_v)G_v(P_2,m_v) \nonumber\\
&+& \frac{c_4-c_3}{2}(G_v(P_1,m_v)+G_v(P_2,m_v)) + 1-c_4 \Big],
\label{eq:F^VMDp}
\end{eqnarray}
with mass dependent couplings, $c_3 = 2-c_4^0 + c_3^m m_\pi^2$ and $c_4 = c_4^0 + c_4^m m_\pi^2$.
In this fit form, the propagator of the excited vector meson is approximated by a constant 
$G_{v'}(P,m_{v'})\sim 1$ (see Fig. \ref{diagram2}). 
We have four free parameters. 
This parametrization~\cite{Yamawaki:2003} satisfies
the  anomaly relation (\ref{eq:f_p0}). 
We impose a constraint $\lim_{m_\pi^2\rightarrow 0}(c_3+c_4)=2$, so that the
transition form factor in the high $Q^2$ limit scales as $1/Q^2$ as suggested by perturbative QCD~\cite{Lepage:1980}. 
We note that a phenomenological estimate leads to $(c_3+c_4)/2|_{\rm phenom}=1.06(13)$ \cite{Yamawaki:2003}.  
Setting the fit range to $0.039\le(aP_2)^2\le 0.347$, we obtain the dashed curves in Fig. \ref{fig:F^lat}. 
Compared with the naive VMD fit (\ref{eq:F^VMD}), the
agreement of the fit with the data in momentum range of $(aP_2)^2 >0.039$ is 
improved especially at larger quark mass region. 
Fit parameters obtained using (\ref{eq:F^VMDp}) are $c_4^0=1.20(21)$, 
$c_3^m=-8.4(8.6)$, $c_4^m=-1.3(5.8)$ and $X_a=0.0243(29)$. 
Again $X_a$ is consistent with the expectation.  
Furthermore $c_4^0$ is consistent with unity,
which implies that the VMD approximation is  reasonable and the
coupling with the excited vector meson state gives a sub-leading contribution. 
In Fig. \ref{fig:F^lat} we also show the comparison with VMD and VMD+ curves
in the physical pion and $\rho$ meson mass.  
The difference between the two lines are almost negligible.

The construction of reasonable fit function for $f_{\pi^0\gamma^*\gamma^*}$ 
over higher momentum region is important when estimating the hadronic L-by-L diagram, because 
it is necessary to integrate $f_{\pi^0\gamma^*\gamma^*}(P_1,P_2)$ from zero to infinity in both $P_1^2$ and $P_2^2$. 
This study provides suggestive information that the VMD model describes the data of 
$f_{\pi^0\gamma^*\gamma^*}(P_1,P_2)$ reasonably well, and inclusion of its excited state significantly 
extends the region to larger momentum.


\section{Summary}

We have studied the $\pi^0\rightarrow\gamma\gamma$ transition form factor by calculating 
the three point function, $\langle AVV\rangle$, on the $N_f=2$ dynamical overlap fermion configurations.
The data of $f_{\pi^0\gamma^*\gamma^*}$ is fitted to the VMD motivated functions, and in the chiral and 
the on-shell photon limits the ABJ anomaly, $f_{\pi^0\gamma\gamma}(0,0)=1/(4\pi^2 f_\pi)$, is reproduced.
It is also found that the fit function based on the VMD model well describes the behavior of mass dependence 
at the lowest momentum in which the higher resonance contribution is small. 
Although the present work is feasible study and there are many things
which have to be understood, 
{\it e.g.} the contribution from higher excited state of pion or vector mesons, 
disconnected diagram effects, finite volume and fixed topology effects, 
comparison with a prediction from perturbative QCD, 
our result encourages us to apply to the non-perturbative estimate 
of the hadronic L-by-L scattering amplitude.

\begin{acknowledgments}
Numerical calculations are performed on IBM System Blue Gene Solution
and Hitachi SR11000 at High Energy Accelerator Research Organization (KEK) 
under a support of its Large Scale Simulation Program (No.~09-05).
This work is supported by the Grant-in-Aid of the Japanese Ministry of Education
(No. 19740121, 
     19540286, 
     20105001, 
     20105002, 
     20105003, 
     20340047, 
     21105508, 
     21674002, 
     ).
\end{acknowledgments}

\end{document}